\documentclass[a4paper]{spie} 
\usepackage[]{graphicx}
\usepackage[switch]{lineno}
\usepackage{xcolor} 
\usepackage{url}
\usepackage{amsmath}
\graphicspath{ {figs/} }
\title{Generalized Training for Neural Network Learnability: a Spectral Methods Approach }

\author{Altai Perry and Luat T. Vuong
\skiplinehalf
{\small Department of Mechanical Engineering, University of California at Riverside, Riverside, CA 92521}
}

\authorinfo{E-mails: LuatV@UCR.edu and APerr010@UCR.edu}

\begin{document} 
 \maketitle 

\begin{abstract}

Hybrid optical neural networks (HONNs) offload some electronic computation to optical preprocessors to achieve low-power and fast training and inference phases in machine learning tasks.
Our contribution to the development of HONNs is a spectral-methods paradigm for building synthetic training data for machine-learned models. 
Here, our synthetic training image data does not resemble the image test data. 
As a result, the neural network focuses on learning specific features parameterized by the synthetic training data. 
Within this paradigm, a dataset's singular value decomposition entropy indicates {\it learnability}, i.e., how rapidly a model converges.
Subsequently, we train a neural network model to rapidly learn specific features for further downstream analyses. 
\end{abstract}


\section{Introduction}

Machine-learning neural network systems currently depend on large volumes of training data for various applications, including manufacturing, quality inspection, automated decision-making, and unmanned vehicle control \cite{Gutierrez2021, Le2017, Endres2022, Figueira2022}. 
In all these cases, as machine-learning models further expand in generalization, they require significant volumes of varying training data and increases in compute times.
When there is an insufficient amount of labeled training data, advanced generative models can augment real data by producing synthetic datasets that closely mimic the characteristics of the test data \cite{Endres2022, Figueira2022}. 

As the constraints on synthetic training datasets increase and the computational demands of deep-learning pipelines rise \cite{Thompson2020, Justus2018, Kortylewski_2019_CVPR_Workshops, Jaipuria_2020_CVPR_Workshops, Endres2022}, Hybrid optical neural networks (HONNs) present a promising solution to provide fast, parallelized computing. 
This combination of optical computing and back-end computing is often referred to as hybrid optical electronic neural networks \cite{Muminov2020,Muminov2021,Chang2018,Miscuglio2020} or deep optics\cite{Wetzstein2020}.  
HONNs reduce the time and power requirements for all-electronic computation \cite{Jutamulia1982} while retaining the flexibility of electronic systems. 
Fourier optical preprocessing, where a detector is placed in the Fourier-plane of an image, is a particularly analytically tractable system for image processing. 
In spite of the phase retrieval problem, Fourier optical preprocessing flattens the computation needed to execute similar processes in machine learning \cite{Zhang:14,Aslani2022}. 
However, in most examples to date, the approach taken with optics mimics prior all-electronic pipelines with training data that resemble the test data \cite{Yan2019,Fu2024,Zuo2019}.  
Here, we aim to leverage opportunities posed by Fourier-plane imaging in the stage model training.

\begin{figure*}[t!]
\centering\includegraphics[width=.7\textwidth]{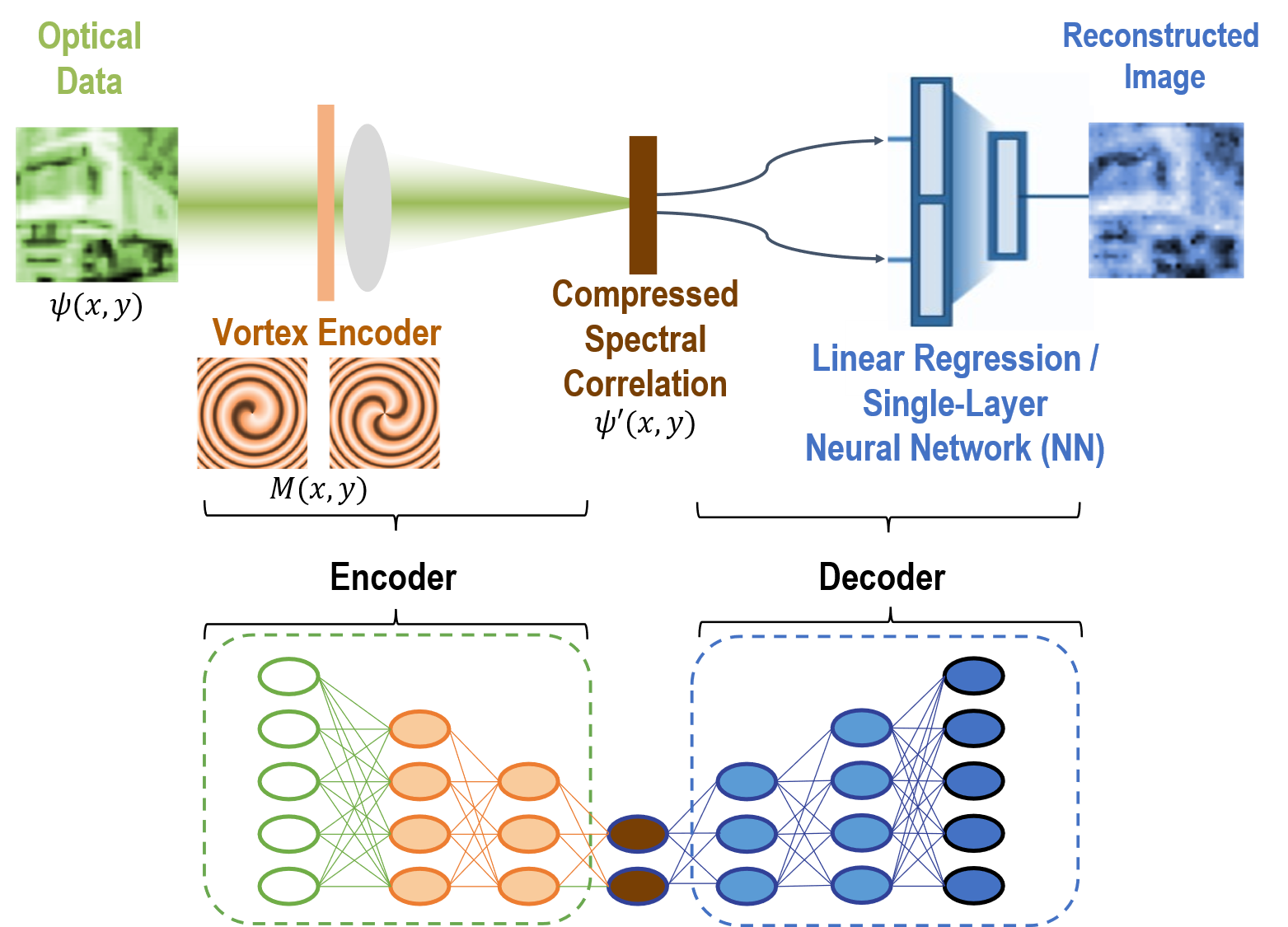}
    \caption{Schematic for a dual vortex, optically encoded and electronically decoded system. A set of spectral correlations of some source image, $\psi(x,y)$, are produced in the sensor plane, $\psi'(x,y)$ (top). The resulting data has some of the original image's phase information. This system is analogous to an autoencoder, machine-learned architecture that distills features and denoises images (below).}
\label{fig:paradigm}
\end{figure*}

To exploit $k$-space correlations, we propose an approach to synthetic training data with a Fourier encoded-aperture optical system where the training images are parameterized with regards to their spectral modal density and do not resemble the test data. 
Our approach advances prior efforts at a generalized of ``Universal Training Dataset'' \cite{Muminov2021} with spectral methods in the training dataset generation.
Spectral methods, employed in numerical simulations and big-data analytics, leverage modal or nonlocal representations of data to identify relationships that may be obscured by the data as a whole \cite{Trefethen,Feit1982,Weie2006, Burns2020}. 
Extended in clustering algorithms, spectral methods underline unsupervised machine learning \cite{Lu2020, Carleo2019}. 
With optics, spectral representations have been leveraged with encoded wave propagation \cite{goodman2005introduction, CANDES2015277, Soltau2022, Williams2006}, and next-generation computational cameras \cite{Novak2020, Deng2020, Antipa:18, Tian:15, flatcam, Sitzmann2018}. 
Spectral methods and representations offer 
compression of vast, multi-dimensional volumes of information into smaller sets of physically interpretable quantities, which reduces the computation cost of deep learning algorithms. 

We evaluate our spectral-methods training paradigm with training sets for our HONN system consisting of speckle images of varying visual density. 
The system in question is represented in Fig. \ref{fig:paradigm} and relies on the use of  dual vortex encoders to rapidly solve the inverse phase problem \cite{Muminovvort}.
Here, we use the dual-vortex diffractive encoder to provide a transformation between sensor and signal images that preserves some phase information \cite{Kularia_2024, Novak2020, Wang2020, Kularia2024}--we solve for this transformation via a single layer NN. 
Our parallel, dual-vortex-encoded aperture system solves the phase retrieval image reconstruction problem as the linear solution to two image equations. 
The process achieves accurate convergence with significantly fewer than 10,000 small training images and, as a result, the inverse transformation between image field and sensor data can be rapidly calculated via linear regression models. 
We underline this important point: this linear solution to the phase retrieval problem---which, with the vortex encoder, becomes an easily learned transformation between the Fourier-plane intensity images and field-encoded images---enables our spectral methods framework.

\begin{figure*}[tbh]
 \centering
 \includegraphics[width=.8\textwidth]{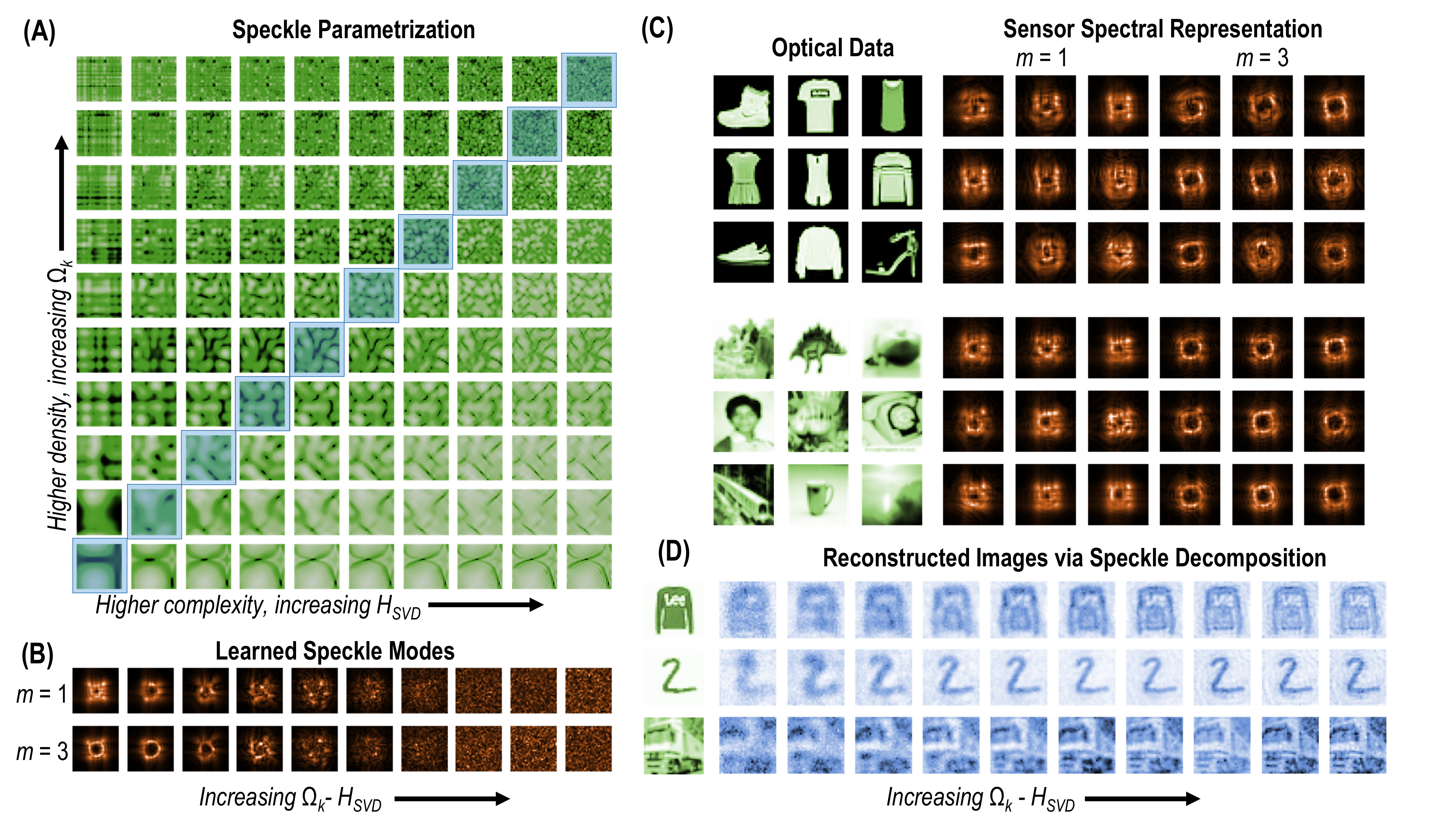}
 \caption{(A) Representative speckle images from 100 training sets and models, which are parameterized with Singular Value Decomposition Entropy ($H_{\rm SVD}$) and Speckle Analogue Density (SAD, $\Omega_k$). The diagonal, blue-highlighted images represent the primary dataset from which the other 90 are formed.
 (B) Representative vortex-encoded sensor patterns applied in parallel with topological charges $m$ = 1, 3 for the primary dataset. 
 (C) Images from the MNIST fashion and CIFAR-100 datasets and their vortex-encoded, sensor patterns for $m=1,3$.
 (D) Representative reconstructed images using the pre-trained models from the primary datasets.
 }
 \label{fig:paradigm2}
\end{figure*}

\section{Methods and Results}

Here, we first describe the HONN setup and explain briefly how it works. We then discuss how we parameterize images based upon `relative span' and `modal density'. This leads to a discussion about the impact of binning our training data via our metrics for span and density for training speed and reconstruction fidelity.

\subsection{Facile Phase Decoding Enabled by a Dual-Vortex Encoder}

Vortex encoded diffraction introduces a mask $M(x,y)$ that shapes the sensed patterns and enables the recovery of phase in the Fourier plane.
In some cases, the encoder may be a metasurface or reconfigurable diffractive optical element \cite{Rodrguez2021, Bernstein2021, Xiang2023}. 
However, in our work, we use optical vortex encoders to modulate the phase of the electric field pertaining to a particular image. This phase modulation takes the form $M_m(x,y) \propto e^{i m \phi}$, where $\phi$ is the azimuthal coordinate in the aperture plane parameterized by the Cartesian coordinates $\left( x,y \right)$.
The integer value $m$ of the imposed vortex is the topological charge that is a measure of the phase helicity--how many times the phase wraps around a singularity given one rotation.

Given the vortex phase mask, $M_m$, the augmented sensor field pattern, ${\psi}_m'(u,v)$, is proportional to the Fourier transform of the field before the lens aperture, $\psi(x,y)$. 
The transform pairs are $(x,y)\leftrightarrow (u,v)$ \cite{goodman2005introduction}.
Given this modulation, the inverse phase problem becomes:
\begin{equation}
\left| \psi_m'\right| \propto |\mathcal{F}\left(M_m \cdot\psi\right)|\longrightarrow \psi, \label{eq: general scheme}
\end{equation} 
where ($\cdot$) denotes the element-wise Hadamard product. 

Given DeMoivre's theorem \cite{Muminovvort}, the vortex encoder produces off-axis field gradient patterns:
\begin{eqnarray}
    \psi'_{\pm m} (u ,v )_{(u,v) \neq (0,0)} &\propto& \mathcal{F}\left\{ M(x,y)\psi(x,y)  \right\}\\
&\approx& \left[ \left( \partial_v - i  \, \partial_u \right) \right]^{|m|} \psi_o'
    \label{eqn: single vortex derivative},
\end{eqnarray}
where $\partial_u$ and $\partial_v$ are the partial derivatives with respect to $u$ and $v$. $u$ and $v$ are the Fourier transform pairs of $x$ and $y$ which relate via $u=x/\lambda f$ and $v=y/\lambda f$. On axis, the sensor pattern provides shape identification associated with $m$ \cite{Vuong2021}.

Solving for the sensor intensity pattern from Eq. \ref{eqn: single vortex derivative}, the resulting profile results in an expression consisting of a sum of polynomial gradients, which we present here in closed form:

\begin{equation}
    \mathcal{X}_{(u,v) \neq (0,0)} = |\psi'_m|^2 _{(u,v) \neq (0,0)}\propto 
     \sum _{\substack{j,k,l_1, l_2\\j+k+l_1+l_2 = |m|}} s_{j,k,l_1,l_2} \,  \\\left[\partial_u^{k+l_1}\partial_v^{j+l_2}\psi_o'\right] \; \left[\partial_u^{k+l_2} \partial_v ^{j+l_1}{\psi_o'}^*\right] \label{eqn:sensor pattern} 
\end{equation}
with linear weighting coefficients
\begin{equation}
    s_{j,k,l_1, l_2}  = \frac{|m|!}{j!\,k!\,l_1!\,l_2!}\,(\pm i)^{l_1+l_2}(-1)^{l_2}.
\end{equation}
Equation \ref{eqn:sensor pattern} identifies the vortex-encoded off-axis sensor pattern as a superposition of gradient correlations of the unencoded Fourier-plane field $\psi_0'$. 

This means that the map between signal fields and sensor intensity patterns is the solution of a 2-{\it D} polynomial regression problem of order $2m$, the solution of which is approximated by simple NNs \cite{GoodBengCour16, Emschwiller2020, Cheng2018}. Notably, Eq. \ref{eqn: single vortex derivative} does not account for the on-axis shape-dependent pixel intensity \cite{Vuong2021}.  

The inverse solution [Eq. \ref{eq: general scheme}] can be accurately estimated by a learned, linear matrix transformation with two or more vortex encoders. 
Two encoders are needed to solve for two variables: phase and amplitude.
The interpolation of $\psi$ involves $\nabla\psi'_0\rightarrow \psi'_o$ and $\psi'_0 \propto \mathcal{F}\psi \rightarrow \psi$. 
In fact, when $m=0, \pm 1$, then the sensor patterns provide a quadratic relation for the gradient of $\psi_o'$, which is accurately solved by a single-layer NN without hidden layers \cite{GoodBengCour16}. 
Higher-order $m$ provide additional terms in Eq. \ref{eqn:sensor pattern} to fit $\nabla\psi_0'$, which yield higher accuracies (as with a Taylor expansion) at the cost of deeper NNs. 
In this analysis, we use $m=\{1,3\}$ to make use of non-redundant symmetry. 
The choice of topological charge will affect reconstruction and classification; however a further analysis of vortex choice is beyond the scope of this text. We turn our attention to the choice of training dataset.

\subsection{Parameterized Training Data}

In the previous section, we explained how the phase recovery problem is tractable with vortex encoders and a small, fully connected neural network.
NNs, especially when shallow, are highly dependent on the quality of their training data. 
In an effort to minimally bias what the NN reconstructs, we parameterize randomly constructed data to study how what spectral features are best learned.
To do this, we first generate ten spectrally-parameterized (via singular value decomposition) speckle training-image datasets. 
These ten datasets are subsequently used to produce 90 additional synthetic training datasets. 
The speckle training patterns are motivated by the Fresnel propagation of the second-order statistics of a randomly scattered beam \cite{Goodman2020}; speckles are a tractable image basis that are observed in a variety of imperfect, coherent-imaging systems \cite{Aime2021, Devaud2021, Goodman1976, Premellieu2024}. 
The datasets are parameterized by SVD entropy ($H_{\rm SVD}$), a measure of relative image span, and Speckle Analogue Density (SAD), a measure of average spatial $k$-vector frequency.  
The relative span is defined by singular value decomposition (SVD) \cite{1102314, BruntonTB, WANG2019423}, which is a ``virtual channel width''---i.e., the information transferred per shot. This virtual channel width is $H_{\rm SVD}$, or SVD entropy $H_{\rm SVD}$:
\begin{eqnarray}
 H_{\rm SVD}&=&\frac{-1}{\log_2 N} \sum^{N}_{i}\tilde{\sigma}_i\log_2 \tilde{\sigma}_i
 \label{eqn: SVDH}\\
 \tilde{\sigma}_i &=& \frac{\sigma_i}{\sum_j^N \sigma_j},
\end{eqnarray}
where $\sigma_i$ is the $i^{th}$ singular value (SV) and $N\times N$ is the dimension of the image. 
Our definition [Eq. \ref{eqn: SVDH}] yields values of $H_{\rm SVD}$ with a range of (0,1] and in contrast to prior work \cite{Alter2000}, we use linear values of $\sigma_i$ in Eq. \ref{eqn: SVDH} to signify a measure of the electric field rather than $\sigma_i^2$ associated with image probabilities \cite{Weng2022}. 
Our measure $H_{\rm SVD}$ is analogous to the 1-{\it D} measure of Shannon information, where the accurate transmission of higher-entropy signals requires more bits \cite{Lapidoth2009}. 
Shannon information does not capture 2-{\it D} image complexity \cite{razlighi2009comparison} since a shuffle of image pixels yields the same Shannon entropy.

$H_{\rm SVD}$ is a relative measure of bandwidth or image span \cite{Perry2022}; larger values of $H_{\rm SVD}$ signify that more eigenimages/spatial modes are needed to approximate an image. 
In other work, $H_{\rm SVD}$ is also a stopping measure or figure of merit for Deep-NN image reconstruction \cite{Buisine2021}, a measure of self-similar fractals \cite{Weng2022}, and a measure of the distribution of electric-field modes in an optical system \cite{Miller:19, Devaud2021}. 

The first ten of the 100 datasets are composed of 10,000 $32 \times 32$ px speckle patterns as training data $\mathbf{\psi}_{\rm train}$ with the generating equation, 
\begin{equation}
    \mathbf{\psi}_{\rm train}^{(x,y,\gamma)}  =\Re\left[ \mathcal{F}^{-1} \left\{ (k_x^2 + k_y^2)\gamma e^{-(k_x^2 + k_y^2)\gamma+ i \mathbf{\Lambda}} \right\} \right],
    \label{Eq.n:data_gen}
\end{equation}
where the spatial frequencies $k_x$ and $k_y$ are limited to the range: $[-k_{\rm max},k_{\rm max})$, $\gamma$ is a density parameter, and $\Lambda$ is a $32 \times 32$ random array with uniform distribution between [0, 2$\pi$]. 

SAD refers to the root mean-squared spatial frequency, $\Omega_k$, or
\begin{eqnarray}
 \Omega_k = \frac{\sqrt { \langle k_x^2\rangle +\langle k_y^2\rangle }}{k_{\rm max}}.
\end{eqnarray}
The values of SAD, $\Omega_k$, like $H_{\rm SVD}$, have a range of (0, 1]. Given the explicit equation for the generated test datasets in Eq. \ref{Eq.n:data_gen}, 
\begin{eqnarray}
\Omega_k = \frac{1}{k_{\rm max}}\sqrt{\frac{2}{\gamma}}.
\end{eqnarray}

In total, 100 NN models are trained with a validation split of 10\%, each from 100, 10k-image datasets [Fig. \ref{fig:paradigm2}(A)]. Test accuracy is verified using mean squared error (MSE) and structure similarity index (SSIM) of four standard test datasets: MNIST Numbers, MNIST Fashion, CIFAR-10, and CIFAR-100. All NN trained use the tensorflow built-in `Adam' optimizer. The ten primary speckle training sets are subsequently used to generate the additional 90 by switching the eigenimages and eigenvalues. 
Each set of images is used to train a no-hidden-layer NN with linear activation and no bias. The resulting model is a linear transformation of the form,
\begin{equation} 
\mathbf{\psi}_{\rm pred}=\mathcal{X}\mathbf{W}^T, \label{Eqn:weights}
\end{equation}
where $\mathbf{\psi}_{\rm pred}$ is the approximated original image, $\mathcal{X}$ is the sensor plane intensity and $\mathbf{W}^T$ are the weights of the derived model. Since we use a single layer NN, these weights are the inverse linearized transfer function that solves the inverse phase problem. 

\subsection{Parameterized Learning} 

Since we have chosen a single-layer, linear NN to carry out the backend reconstruction, we can solve explicitly for the epoch-on-epoch reduction in loss via gradient descent. The derivation starts with the transformation preformed on a data set $\psi$ consisting of $N$ images. The image captured on a camera sensor at the focal plane is given by the intensity pattern of the Fourier transform of the electric field. We denote this electric field as the Hadamard (or element-wise) product of 'mask', $M$, with each image in $\psi$. The resulting image is denoted by $\psi'$

\begin{equation}
\psi' =\left(\mathcal{F}(M\cdot \psi)\right)^*\left(\mathcal{F}(M\cdot \psi)\right),
\end{equation}
where $cdot$ denotes the Hadamard product.

If $\psi'$ is defined as above and ground truth is $\psi$, we can define mean square error loss over $K$ images as at epoch $n$,

\begin{equation}
L_n=\sum_k^K (\psi_k-\psi'_kW_n^T)^T(\psi_k-\psi'_kW_n^T).
\label{eqn:loss}
\end{equation}



Stepping forward, epoch on epoch, through gradient descent yields an epoch-on-epoch loss difference. While stochastic gradient descent, like the Adam optimizer used in this investigation, will adjust learning rate, $\alpha$, to avoid small local minimums, the gradient in loss is largely given by the data manifold itself. The gradient descent algorithm, in general, discovers the next $W$ via

\begin{equation}
W_{n+1}=W_n-\alpha\frac{\partial L_n}{\partial W_n}.
\label{eqn:gradient_descent}
\end{equation}

Substituting Eq. \ref{eqn:loss} into Eq. \ref{eqn:gradient_descent} yields

\begin{equation}
W_{n+1}={W_n}+2\alpha \frac{1}{K} \sum^K_k \left({\psi'_k}^T{\psi'_k} {W_n}^T- {\psi_k}^T{\psi_k} \right),
\end{equation}

\begin{multline}
\Delta L_k=
{\psi'_k}^T{\psi'_k} {W_{n+1}} ^T - {W_{n+1}}{\psi'_k}^T{\psi'_k}+ { W_{n+1}} {\psi'_k}^T{\psi'_k} { W_{n+1}} ^T  - W_{n}{\psi'_k}^T{\psi'_k} { W_{n}} ^T + {\psi'_k}^T {\psi'_k}W_n^T + W_n{\psi'_k}^T {\psi'_k}.
\end{multline}

Since $W_{n+1}=W_n+\Delta W_n$, the difference between terms with $W_{n+1}$ and $W_n$ leaves us with only terms relating to $\Delta W_n$. This means that every term has at least a leading coefficient of $\frac{2\alpha}{K}$. The terms that have $W_{n+1}^2$ are nearly zero. That leaves us with the expression:

\begin{equation}
\Delta L_{n,k} \frac{K}{2\alpha} = \psi^{\prime}_k W_n^T \left( W_n\psi_k^{\prime T} - \\ \psi_k^T \right)\psi_k^{\prime} \psi_k^{\prime T} +W_n\psi^{\prime T}_k \psi^{\prime}_k\psi^{\prime T}_k \left( \psi^{\prime}_k W_n^{T} -\psi_k \right) ,
\end{equation}

or, more succinctly with an error term, $\epsilon_{n,k}$, defined as $\left( \psi'_kW^T_n-\psi_k \right)$:

\begin{equation}
\Delta L_{n,k} \frac{K}{2\alpha} = \psi^{\prime}_kW_n^T \epsilon_{n,k}\psi^{\prime}_k\psi_k^{\prime T} +W_n\psi_k^{\prime T} \psi^{\prime}_k\psi_k^{\prime T} \epsilon_{n,k} .
\label{eqn:training_time}
\end{equation}

The loss-on-loss change is proportional to the error, $\epsilon_{n,k}$. This error is smaller when both the sensor-plane image and the ground truth have larger relative spans.

\begin{figure}
    \centering
    \includegraphics[width=0.8\linewidth]{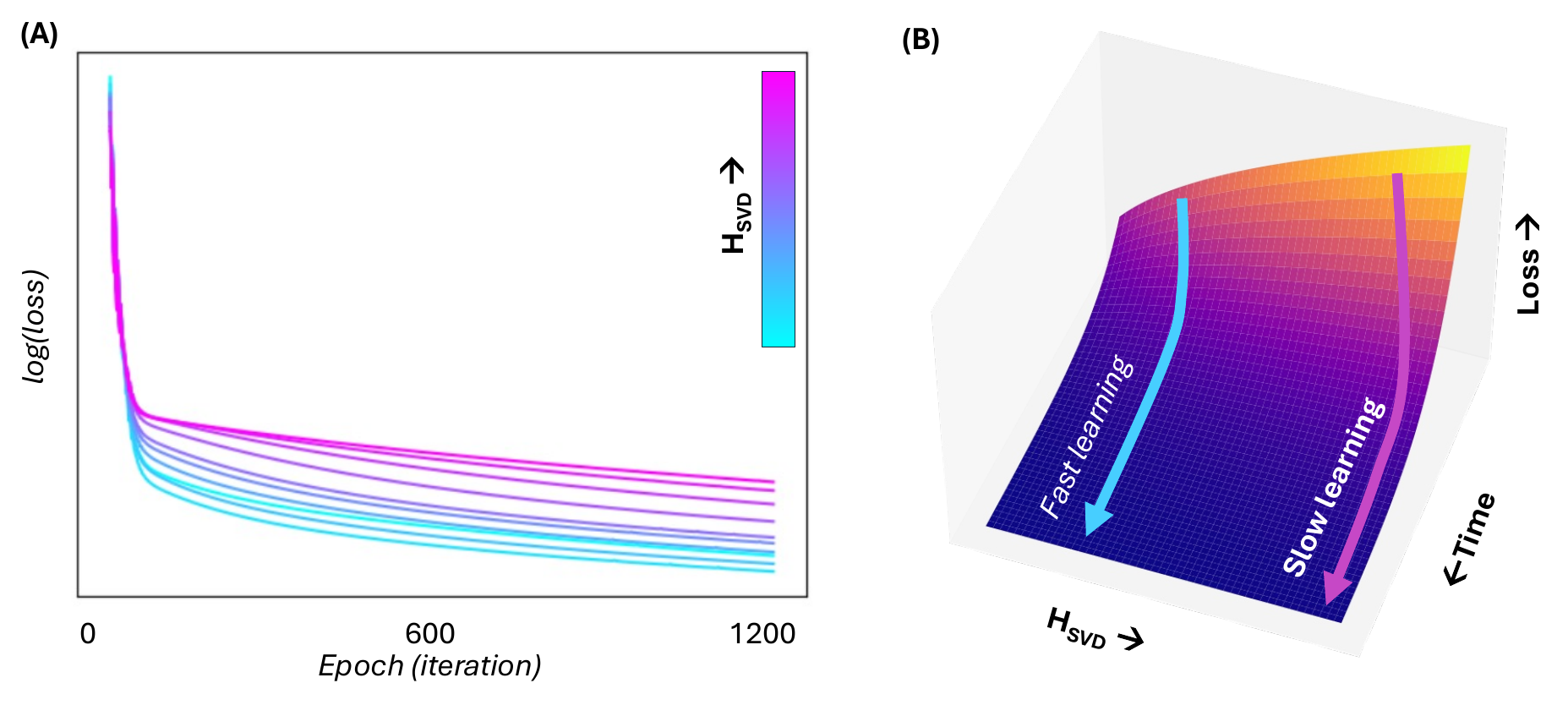}
    \caption{(A) Several machine learning convergence plots are shown on a log-scale. Each of these convergence curves correspond to a model trained on a dataset with a different $H_{SVD}$. (B) Those same plots are extended to reflect the average $H_{SVD}$ of the images used to train the models. Given this view, it is clear that models with lower $H_{SVD}$ train faster--reach training loss-convergence.}
    \label{fig:isochrones}
\end{figure}

The NNs trained with higher average $H_{\rm SVD}$ images take longer to converge. 
We preform the training with randomly generated training sets of similar $H_{\rm SVD}$ and SAD ($\Omega_k$) statistics and run experiments in triplicate. 
Figure \ref{fig:isochrones}(A) shows the initial ten models' MSE at four slices of time. Each model's MSE is plotted against the average $H_{\rm SVD}$ of the model's training data. 
Across randomized triplicate training sets, the measured accuracy at analyzed `stopping points' has a variance of less than $10^{-5}$ per pixel, which indicates that the epoch-on-epoch convergence profiles are reproducible [Fig. \ref{fig:paradigm2}]. 

To explain the relationship between $H_{\rm SVD}$ and convergence time, we calculate the epoch-on-epoch $\Delta L$, the epoch-by-epoch change in loss through gradient descent. The derivation is provided in the appendix. 
The epoch-by-epoch change in loss is proportional to the error, $\epsilon_{n,k}$ and is smaller for matrices with a slow SV drop off or high $H_{\rm SVD}$ and large relative span. 
A higher convergence time is attributed to lower epoch-to-epoch change in loss governed by the epoch-to-epoch error. 
Images whose eigenimages are weighted more evenly and carry higher $H_{\rm SVD}$ are less easily learned. 
This trend is dependent on the $H_{\rm SVD}$ regardless of the dataset eigenimages [Fig. \ref{fig:isochrones}(B)]. 
High $H_{\rm SVD}$ training data yields lower NN learning rates regardless of the SAD ($\Omega_k$). 

\subsection{Reconstruction}

\begin{figure*}[t]
 \centering
 \includegraphics[width=.8\textwidth]{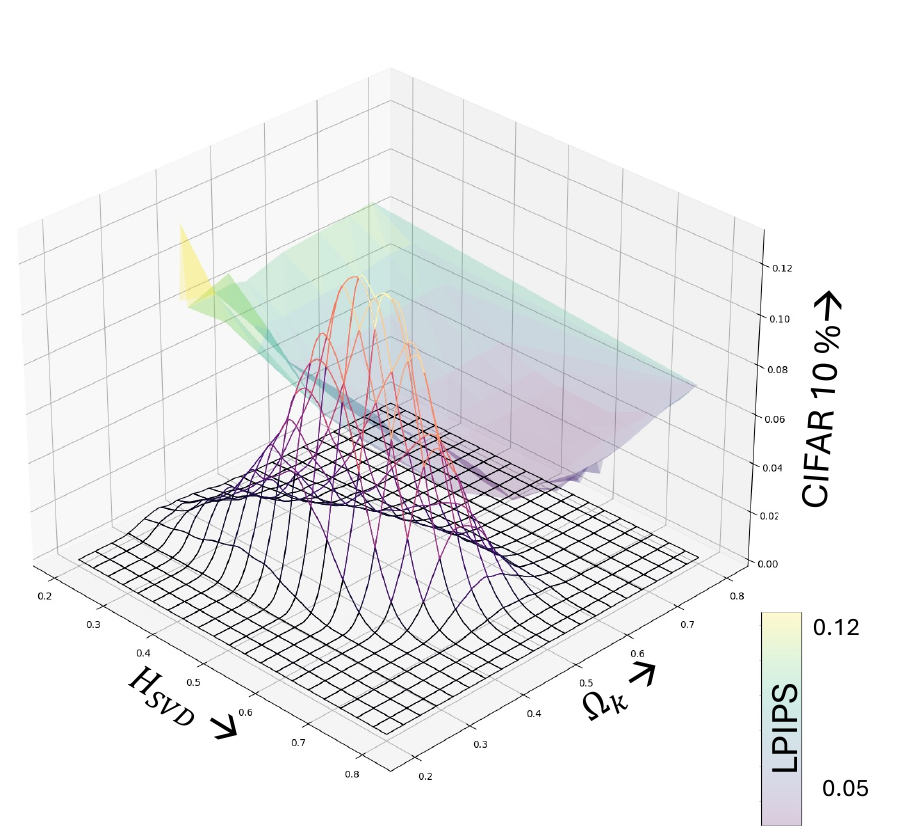}
 \caption{The LPIPS of models trained on the synthetic training datasets as described earlier in this text is plotted against the average SAD ($\Omega_k$) and $H_{SVD}$. Lower (darker) LPIPS refers to greater fidelity to the original image. Underneath is a density map of the CIFAR10 dataset also parameterized by SAD and $H_{SVD}$. The lowest LPIPS value (the highest fidelity) coincides with the height of CIFAR10 density.}
 \label{fig:oyster}
\end{figure*}

In contrast, SAD parity is a heuristic for structurally similar reconstruction. SAD parity refers to similar $\Omega_k$ in the train and test data.
From each of the models visualized in Fig. \ref{fig:paradigm2}(B), we calculate the SSIM with the CIFAR-100 dataset. 
The colored dots show the average SSIM while background contour shows the histogram of the CIFAR-100 test dataset's $H_{\rm SVD}$ and $\Omega_k$ [Fig. \ref{fig:oyster}]. 
While the $H_{\rm SVD}$ has a direct impact on training time, it has little-to-no impact on reconstruction accuracy; instead, SAD parity between training and test datasets dictates reconstruction accuracy. 

The peak in SSIM vs. SAD on the right of Fig. \ref{fig:oyster} is not symmetric: models produced with lower-SAD training images have higher SSIM than the models produced with higher-SAD training images. 
This is largely a result of the SSIM metric: smooth, shape-aligned images have higher SSIM \cite{Wang2004}. 
High SAD would indicate that lower frequency training data is not necessarily included in the training set. 
A balance between high and low regions of SAD are necessary to reconstruct most images. 
The images generated by models with high SAD training data are more visually-dense and are penalized by SSIM. 
Not surprisingly, the models that are trained with similar-SAD images also generate images with low MSE.
(Similar results are achieved with the MNIST Digit, Fashion, and CIFAR-10 dataset.) 

\begin{figure}
    \centering
    \includegraphics[width=0.9\linewidth]{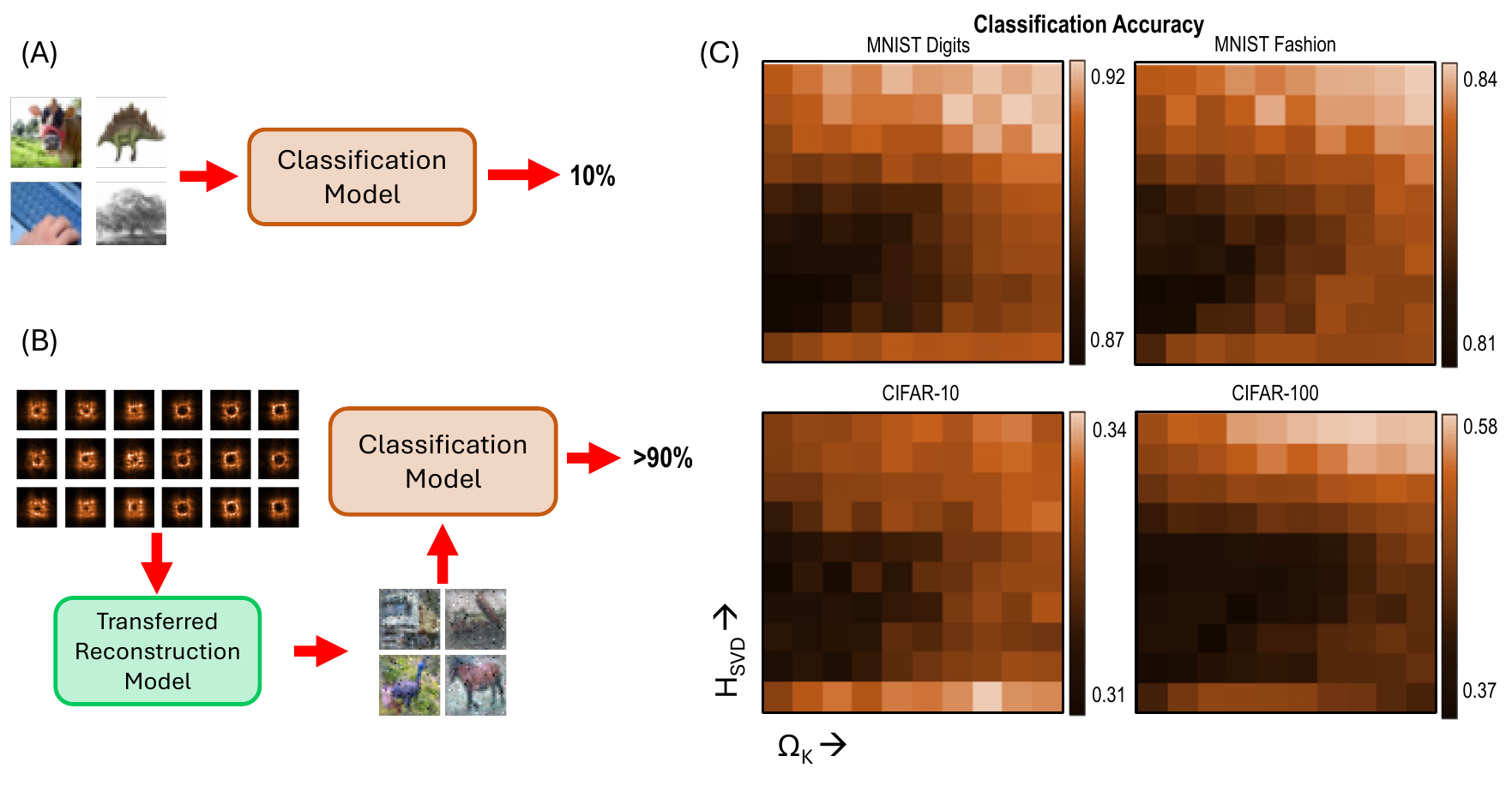}
    \caption{(A) A previously untrained 2-layer dense neural network is used to classify CIFAR-10. Predictably, the classification accuracy of the model is 10\%, which is no better than random chance. (B) Whereas, the same original information can be first preprocessed by a neural network to create images that, when used to train a classification neural network, yield upwards of 90\% accuracy. (C) Compressed representations of the overall landscape of the performance of test data `distilled' with models of varying $H_{SVD}$ and $\Omega_k$. For the most part, high entropy/high SAD data result in better classification performance, regardless of initial reconstruction accuracy.}
    \label{fig:placeholder}
\end{figure}

LPIPS determines patch similarity based upon a pretrained convolutional model, however, another method to determine reconstruction model efficacy for machine vision is to determine potential classification accuracy. As a result, we optically transform and then reconstruct with each of the 100 pretrained reconstruction NNs one of 4 standard, small size datasets: MNIST Digits, MNIST Fashion, CIFAR-10, and CIFAR-100 (Fig. \ref{fig:placeholder}). These reconstructed images are then each used to train a downstream classifier comprising of 2 dense, fully connected layers. These 400 models are evaluated for accuracy and judged against that same neural network architecture trained on the original images. Aside from the MNIST digits dataset, which is well classified by even non-supervised methods, the initial NN architecture is insufficient to classify any image with any more accuracy than chance ($10\%$ for MNIST digits, MNIST fashion, and CIFAR-10; $1\%$ for CIFAR-100). However, there exists some choice speckle-pretrained reconstruction that works as an effective preprocessor to further distinguish classes of the initial dataset. While the accuracy of the resulting models do not contend with the state of the art, the fact that the classification model based upon pre-distilled images can preform better than images that were never augmented through a lossy Fourier-domain transformation is remarkable. We believe that further investigation into semi-random training sets for spectral reconstruction can prove to be instrumental in reducing computational overhead for future machine vision classification. 

\section{Discussion and Conclusions}

The training and learning of neural networks are highly dependent on the training data, in essence--the model will have certain characteristics based upon the `lesson plan' it is given. 
We can design our training datasets with an analogy in cognitive psychology, the `mental load' imparted by different methods of teaching. 
The framework is called cognitive load theory (CLT) and has been used in building teaching plans that most effectively make use of limited temporal cognitive capacity. 
The theory largely parameterizes cognitive load into three separate categories: Intrinsic Cognitive Load (ICL) which relates to the number of pieces of information that must be juggled before understanding of the task is complete, Extraneous Cognitive Load (ECL): pieces of information not relevant to the current task, and Germane Cognitive Load (GCL), which are the necessary pieces of information to communicate.
Given finite load capacity at any given time, the goal is to reduce the intrinsic load, eliminate any extraneous load and emphasize any germane load. 
To translate this into machine learning deliverables we
\begin{enumerate}
    \item quantify the number of pieces of information the model learns (ICL) via $H_{SVD}$,
    \item eliminate extraneous information (ECL) by training with a synthetic dataset that has no additional features, and 
    \item determine what training-set qualities are and are not important for a given task (GCL) via a frequency-based metric, SAD.
\end{enumerate}
Our results also have a tidy analogy to learning in humans: when students are learning a difficult concept---not simply memorizing but learning to achieve mastery for its general applicability---it is valuable to teach the concept with multiple representations \cite{ainsworth2008educational}.
While this approach takes longer, the more time students spend learning, the better they are at recalling that concept \cite{Mulder1986}. 
In our effort, multiple-representation training is analogous to higher-$H_{SVD}$ images: longer convergence times achieve pre-trained models whose post-reconstructed images are more accurately classified.

For {\it machine}-learning, our research primarily tackles model learning rate, model convergence, and reconstruction accuracy \cite{SHIOYA200688, Krizhevsky09learningmultiple, Antun2020, hinton2015distilling}. 
We have discovered that, rather than calculating the image priors via optimization or learning \textit{local} features \`a la convolutional NN, we can instead guess at what spectral or \textit{global} features exist in a class of images and encode them in a generalized synthetic training dataset {\it a priori}. 
With Fourier-domain HONNs the local features are in the training set and its global representation are the sensor images that go to the NN.
The strength and potential drawback of HONN systems lies in the simultaneous and somewhat inseparable operations of both convolutions and filters, which we mitigate with careful parameterization of the training data.
The advantage of our physically realized system is that it is tractable, simple, carries low computational overhead but mirrors and expands on successful computer vision algorithms, namely others that take inspiration from natural processes that build models based upon diffuse correlations \cite{Bi2024,cai24,yan20,Guo2023,shi23}.  

In conclusion, we create a spectral-methods approach to develop training data and to analyze the learning transferred.
As our work has shown, there are trade-offs, since a higher-span dataset as characterized by higher $H_{\rm SVD}$ generally exhibits a reduced learning rate. However, this carries strong analogies to teaching and learning in humans \cite{ainsworth2008educational, Mulder1986}. Extended further, our work may contribute to a unified system of design that also integrates how one recognizes features to the statistical-weighted exposure to training data.

\section*{Acknowledgments}

The authors recognize funding from DARPA: DSO Grant \#D19AP00036 .

\section{Disclosure}
The authors declare no conflicts of interest.

\bibliography{report} 
\bibliographystyle{spiebib} 

\end{document}